\documentclass[preprint,12pt,authoryear]{elsarticle}
\usepackage{amsmath}
\usepackage{amssymb}
\usepackage{amsthm}
\usepackage{mathtools}
\usepackage{multirow}
\usepackage{wrapfig}
\usepackage{subcaption}
\usepackage{array,multirow}
\usepackage{incgraph}
\usepackage{geometry}
\usepackage{ulem}
\usepackage{lineno}

\usepackage{color}

\usepackage{tikz}
\usetikzlibrary[topaths]
\usetikzlibrary{arrows}
\usetikzlibrary{shapes.misc, positioning, arrows, decorations.text, shapes, mindmap,trees}

\journal{arXiv}
\begin{document}
\begin{frontmatter}

\title{How to perform modeling with independent and preferential data jointly?}

\author[labelaff1]{Mario Figueira}
\author[labelaff1]{David Conesa}
\author[labelaff1]{Antonio López-Quílez}
\author[labelaff2]{Iosu Paradinas}
\affiliation[labelaff1]{organization={Universitat de València}, country={Spain.}}
\affiliation[labelaff2]{organization={AZTI, Txatxarramendi Ugartea z/g, 
48395 Sukarrieta}, country={Spain}}

\begin{abstract}
    Continuous space species distribution models (SDMs) have a long-standing history as a valuable tool in ecological statistical analysis. Geostatistical and preferential models are both common models in ecology. Geostatistical models are employed when the process under study is independent of the sampling locations, while preferential models are employed when sampling locations are dependent on the process under study. But, what if we have both types of data collectd over the same process? Can we combine them? If so, how should we combine them? This study investigated the suitability of both geostatistical and preferential models, as well as a mixture model that accounts for the different sampling schemes. Results suggest that in general the preferential and mixture models have satisfactory and close results in most cases, while the geostatistical models presents systematically worse estimates at higher spatial complexity, smaller number of samples and lower proportion of completely random samples. 
\end{abstract}


\end{frontmatter}

% \linenumbers

\section{Modelization}

Continuous space autocorrelation models are widely applied for inferring, predicting, and projecting species distributions \citep{HabitatSuitability_Guisan, SpatialEcologyConservationModelling_Fletcher, Joint_Species_Modelling_Ovaskainen}. Geostatistical models refer to models applied over point referenced data \citep{Geostatistical_Diggle}, while preferential models refer to models that account for the dependence between the process under study and the sampling locations \citep{GeostatisticalPreferentialSampling_Diggle}. Different studies have demonstrated the importance of accounting for the sampling process to infer good abundance estimates \citep{pennino2019accounting, Preferential_Watson, Preferential_Dean}. But, what if we have both randomly and preferentially sampled data? Could we merge them together? If so, how should we model them? Do we need to account for the different sampling schemes?

\begin{figure}[h!]\centering
\begin{tikzpicture}[
roundnode/.style={circle, draw=green!60, fill=green!5, very thick, minimum size=7mm},
squarednode/.style={rectangle, draw=red!60, fill=red!5, very thick, minimum size=5mm},
]
%Nodes
\node[align=center,font = {\bfseries}, xshift=-5.5cm,yshift=0cm, rounded corners] at (current page.center) [rectangle,draw] (Geo) {Geostatistical};
\node[align=center,font = {\bfseries}, xshift=-1.9cm,yshift=0cm, rounded corners] at (current page.center) [rectangle,draw] (Pref) {Preferential};
\node[align=center,font = {\bfseries}, xshift=3cm,yshift=0cm, rounded corners] at (current page.center) [rectangle,draw] (Mix) {Mixture};

\node[align=center,font = {\bfseries}, xshift=-8cm,yshift=2cm] at (current page.center) [rectangle] [rectangle, text width=25mm] (MarkTitle) {Process of\\the marks};
\node [align=center,xshift=-1.9cm,yshift=2cm,rounded corners] at (current page.center) [rectangle,draw,fill=orange!20,line width=10mm,thick,text width=70mm,text height=12mm](frameMark){};
\node[align=center,font = {\bfseries}, xshift=-1.9cm,yshift=2cm] at (current page.center) [rectangle] [rectangle, text width=80mm] (MarkForm) {$y\sim f(y_i|\boldsymbol\theta,\boldsymbol\psi)\quad \& \quad g(\mu_i)=\eta_i + u_i$};

\node[align=center,font = {\bfseries}, xshift=-8cm,yshift=-2cm] at (current page.center) [rectangle] [rectangle, text width=20mm] (PointTitle) {Point \\ Process};
\node [align=center,xshift=-5.5cm,yshift=-2cm,rounded corners] at (current page.center) [rectangle,draw,fill=orange!20,line width=10mm,thick,text width=25mm,text height=12mm](framePoint1){};
\node[align=center,font = {\bfseries}, xshift=-5.5cm,yshift=-2cm] at (current page.center) [rectangle] [rectangle, text width=25mm] (PointForm1) {$\log(\lambda_i)=\beta$};

\node [align=center,xshift=-1.9cm,yshift=-2cm,rounded corners] at (current page.center) [rectangle,draw,fill=orange!20,line width=10mm,thick,text width=34mm,text height=12mm](framePoint2){};
\node[align=center,font = {\bfseries}, xshift=-1.9cm,yshift=-2cm] at (current page.center) [rectangle] [rectangle, text width=34mm] (PointForm2) {$\log(\lambda_i)={\eta_i}^* + \alpha \cdot u_i$};

\node [align=center,xshift=3cm,yshift=-2cm,rounded corners] at (current page.center) [rectangle,draw,fill=orange!20,line width=10mm,thick,text width=51mm,text height=12mm](framePoint3){};
\node[align=center,font = {\bfseries}, xshift=3cm,yshift=-2cm] at (current page.center) [rectangle] [rectangle, text width=51mm] (PointForm3) {$\log(\lambda_i)=a_i[{\eta_i^*} + \alpha \cdot u_i] + \overline{a_i} \cdot \beta$};

\node [align=center,font={\scriptsize\bfseries},xshift=-1.9cm,yshift=-4cm] at (current page.center) [circle,draw](Sum){$\textbf{+}$};

%Lines
\draw[-to] (Geo) to [out=90,in=180,looseness=2] (frameMark);
\draw[-to] (Pref) to [out=90,in=270,looseness=0] (frameMark);
\draw[-to] (Mix) to [out=90,in=0,looseness=2] (frameMark);

\draw[-to] (Geo) to [out=-90,in=90,looseness=2] (framePoint1);
\draw[-to] (Pref) to [out=-90,in=90,looseness=0] (framePoint2);
\draw[-to] (Mix) to [out=-90,in=90,looseness=2] (framePoint3);

\draw[draw, dotted, -to] (framePoint1) to [out=-90,in=180,looseness=1] (Sum);
\draw[draw, dotted, -to] (framePoint2) to [out=-90,in=90,looseness=0] (Sum);
\draw[draw, dotted, -to] (Sum) to [out=0,in=270,looseness=0.85] (framePoint3);

\end{tikzpicture}
\caption{The structure of the three models is identical for the mark process. However, for the point process layer, we observe that the geostatistical model corresponds to a homogeneous intensity, the preferential model follows a log-Gaussian Cox process (LGCP) structure, and the mixture model exhibits a combination of the other two structures precisely. In the case of the mixture model, the parameter $a_i$ indicates whether the $i$-th data point comes from a preferred sample ($a_i=1$) or not ($a_i=0$). The complement of $a_i$ is represented as $\overline{a_i}$.}
\label{fig:ModelStructures}
\end{figure}
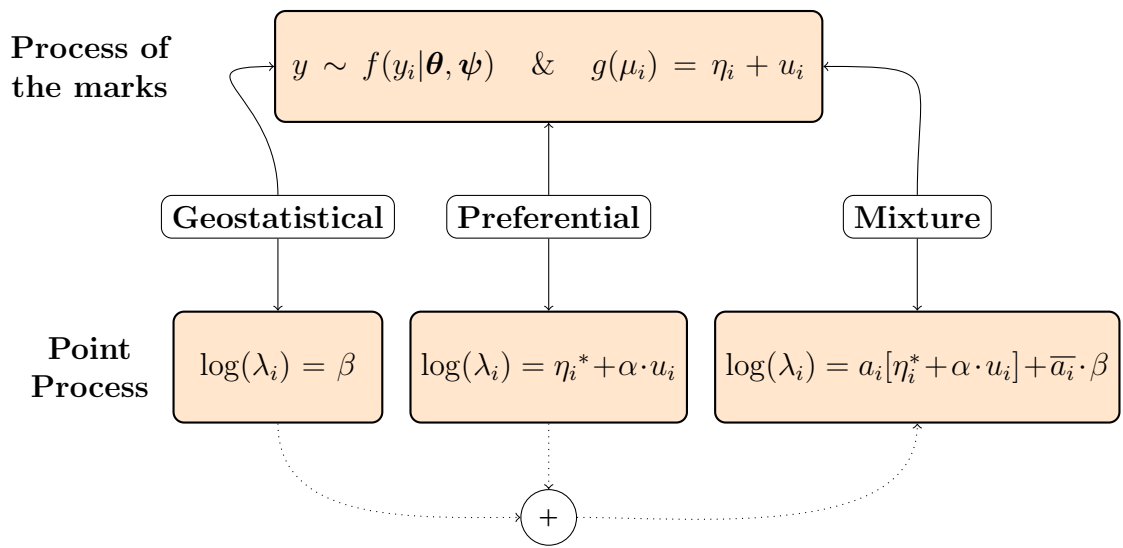

We tested three models as exposed in Fig~\ref{fig:ModelStructures}, the aforementioned geostatistical and preferential models as well as a model encompassing a mixture of both sampling approaches (mixture model from now on). A priori, the more complex mixture model is deemed to be the most appropriate model as it models each data source according to its sampling approach, but in order to check that out, we conducted a simulation study with the final aim of comparing the suitability of all three models.

\section{Simulation Study}

The simulation was designed to test the three different models under different levels of spatial complexity, sample sizes and proportions of randomly and preferentialy drawed samples. Underlying species abundance distributions were based on a geostatistical process, thus spatial complexity was controlled by varying the range parameter. Each scenario was then replicated four times.

Models were fitted using the integrated nested Laplace approximation (INLA) methodology \citep{Rue2009, ReviewINLA_2017}. Model performance was assessed based on three metrics: the Watanabe-Akaike Information Criterion (WAIC) \citep{WAIC_Criterion} to evaluate the goodness-of-fit; the root mean squared error (RMSE) to evaluate the out of sample predictive capacity; and the ratio between the estimated abundance integration and the simulated overall abundance. Results showed no distinct advantage in terms of WAIC. Out-of-sample predictions however highlighted the overestimation tendency of geostatistical models \citep{pennino2019accounting} with lower proportions of random samples and higher complexities of the underlying spatial pattern (see Tables~\ref{tab:RMSEPrediction} and \ref{tab:PropTotalAbundancePrediction}). Interestingly, results showed that the preferential model performed as well as the mixture model even at high proportions of independent samples. Such good performance is driven by the fact that a combination of homogeneous and inhomogeneous point patterns results in another inhomogeneous point pattern with similar shape. Moreover, if the proportion of preferential samples were negligible, the sharing effect between the point process and the geostatistical process in the LGCP model would tend to zero, and therefore two independent processes.

\begin{table}[h!]
    \begin{minipage}{0.475\linewidth}
        \begin{tabular}{|c|c|c|c|} \hline
            & Geo/Pref & Geo/Mix & Pref/Mix \\ \hline
           \textbf{(A)} & \multicolumn{3}{c|}{Spatial range} \\ \hline
           $0.2$ & $1.075$ & $1.083$ & $1.008$ \\ \hline
           $0.5$ & $1.035$ & $1.036$ & $1.001$ \\ \hline
           $0.8$ & $1.021$ & $1.024$ & $1.002$ \\ \hline \hline 
           \textbf{(B)} & \multicolumn{3}{c|}{Proportion of random samples} \\ \hline
           $0.10$ & $1.082$ & $1.079$ & $0.998$ \\ \hline
           $0.25$ & $1.088$ & $1.090$ & $1.002$ \\ \hline
           $0.50$ & $1.035$ & $1.044$ & $1.010$ \\ \hline
           $0.75$ & $1.019$ & $1.026$ & $1.007$ \\ \hline
           $0.90$ & $0.997$ & $0.999$ & $1.003$ \\ \hline \hline 
           \textbf{(C)} & \multicolumn{3}{c|}{Total amount of samples} \\ \hline
           $60$ & $1.061$ & $1.064$ & $1.005$ \\ \hline
           $100$ & $1.042$ & $1.051$ & $1.009$ \\ \hline
           $160$ & $1.037$ & $1.038$ & $1.001$ \\ \hline
           $200$ & $1.037$ & $1.038$ & $1.001$ \\ \hline
        \end{tabular}
        \caption{Mean RMSE proportions were computed for each model and for each value of \textbf{(A)} the spatial range, \textbf{(B)} the proportion of completely random samples and \textbf{(C)} the total amount of samples in the dataset.}
        \label{tab:RMSEPrediction}
    \end{minipage}%
    \hfill%
    \begin{minipage}{0.475\linewidth}
        \begin{tabular}{|c|c|c|c|} \hline
            & Geo/Sim & Pref/Sim & Mix/Sim \\ \hline
           \textbf{(A)} & \multicolumn{3}{c|}{Spatial range} \\ \hline
           $0.2$ & $1.056$ & $0.990$ & $1.001$ \\ \hline
           $0.5$ & $1.017$ & $0.998$ & $1.002$ \\ \hline
           $0.8$ & $1.008$ & $1.000$ & $1.001$ \\ \hline \hline 
           \textbf{(B)} & \multicolumn{3}{c|}{Proportion of random samples} \\ \hline
           $0.10$ & $1.046$ & $0.998$ & $1.004$ \\ \hline
           $0.25$ & $1.041$ & $0.995$ & $1.003$ \\ \hline
           $0.50$ & $1.025$ & $0.987$ & $0.999$ \\ \hline
           $0.75$ & $1.020$ & $1.001$ & $1.004$ \\ \hline
           $0.90$ & $1.002$ & $0.997$ & $0.997$ \\ \hline \hline 
           \textbf{(C)} & \multicolumn{3}{c|}{Total amount of samples} \\ \hline
           $60$ & $1.044$ & $0.997$ & $1.009$ \\ \hline
           $100$ & $1.027$ & $0.993$ & $0.998$ \\ \hline
           $160$ & $1.019$ & $0.994$ & $0.998$ \\ \hline
           $200$ & $1.018$ & $0.998$ & $1.001$ \\ \hline
        \end{tabular}
        \caption{Mean ratio of the total abundance considering \textbf{(A)} different values for the spatial range, \textbf{(B)} the proportion of completely random samples and \textbf{(C)} the total amount of samples in the dataset.}
        \label{tab:PropTotalAbundancePrediction}
    \end{minipage}
\end{table}

\section{Discussion}

Spatial autocorrelation models have a longstanding history of successful application in quantitative ecology. Geostatistical and preferential models have become two fundamental approaches for dealing with point referenced data, the latter when the sample distribution is dependent on the process under study. The aim of this study was to test the suitability of these models as well as a mixture model when the sample includes both randomly and preferentially sampled data.

\begin{figure}[h!]
    \centering
    \begin{subfigure}{\linewidth}
        \centering
        \includegraphics[width=.9\linewidth, height=0.45\textheight]{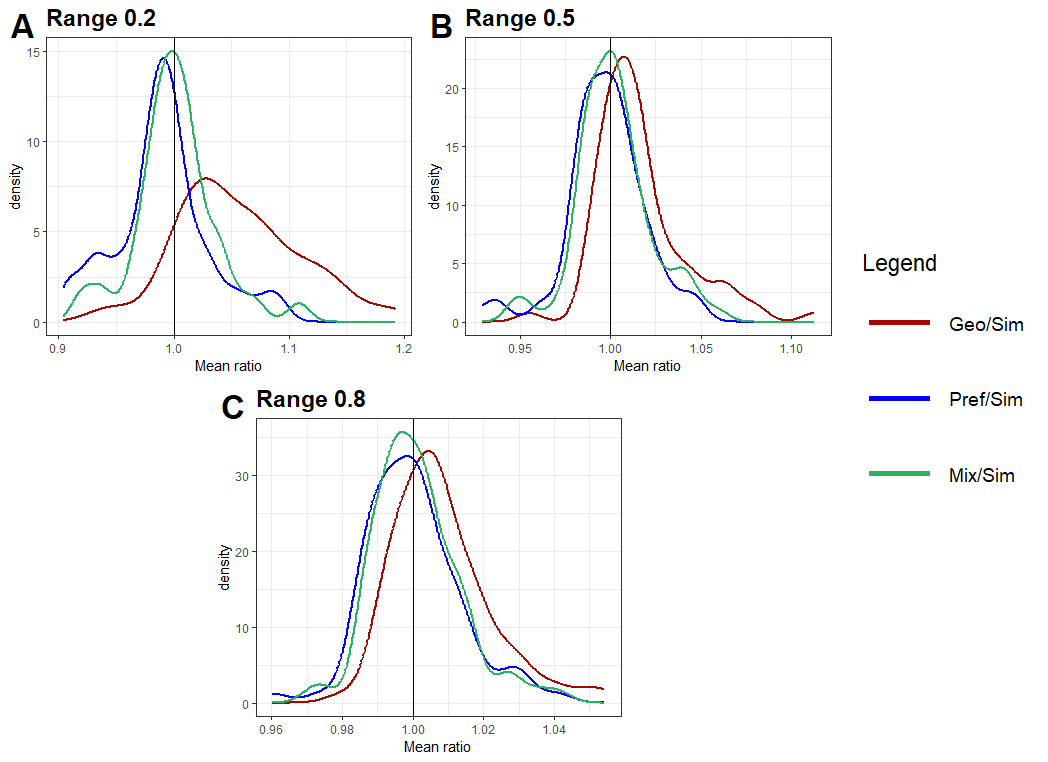}
        \caption{Distribution of the ratios for each model according to range.}
        \label{fig:RangeTotalAbundance}
    \end{subfigure}%
    \vfill%
    \begin{subfigure}{\linewidth}
        \centering
        \includegraphics[width=.9\linewidth, height=0.45\textheight]{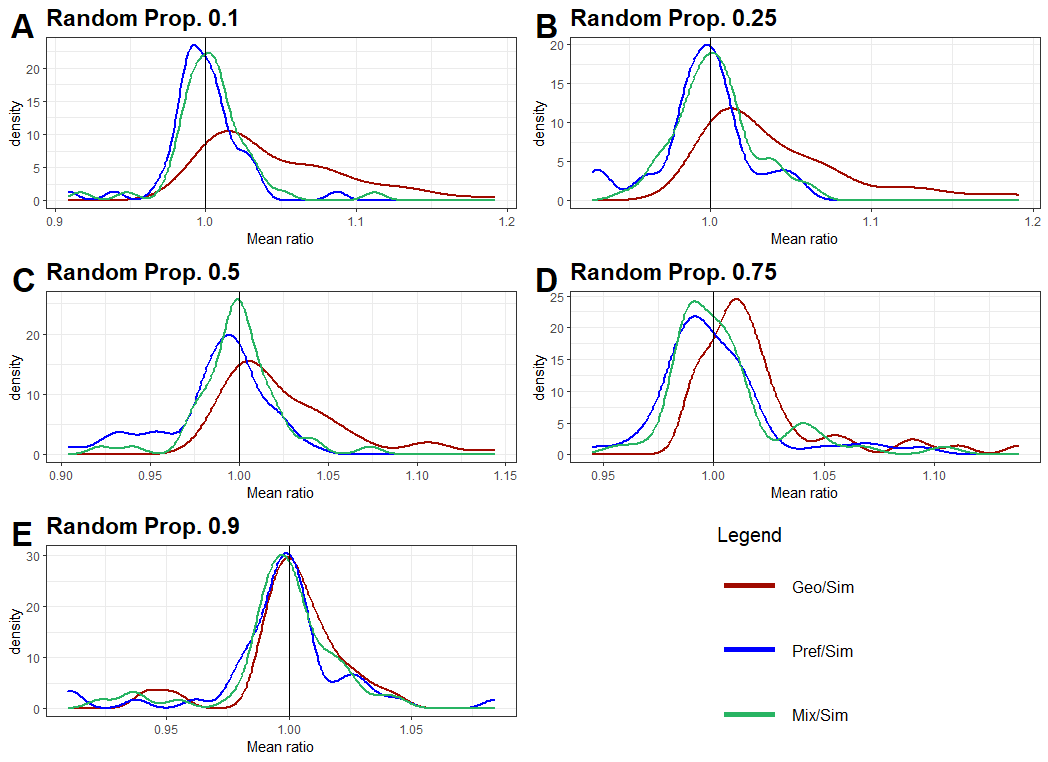}
        \caption{Distribution of the ratios for each model according to the proportion of random samples.}
        \label{fig:RandomPropTotalAbundance}
    \end{subfigure}
    \caption{Distribution of the ratios for each scenario by model, according to the range and proportion of independent samples.}
    \label{fig:RangeRandomPropTotalAbundance}
\end{figure}

The geostatistical model tends to overestimate abundances \citep{pennino2019accounting} when the number of random samples is low and the underlying spatial pattern is complex (see Fig. \ref{fig:RangeRandomPropTotalAbundance}). Given that one can hardly ever know the complexity of the spatial pattern before conducting the study, we discourage the use of geostatistical models unless we have a large proportion of independent samples.

The preferential model perfoms just as well the more complex mixture model, which a priori is the gold standard model given that it accounts for the different sampling processes. Fig. \ref{fig:RangeRandomPropTotalAbundance} shows the similarity of the distributions of the ratios for the preferential and mixture model, and the systematic deviation for the geostatistical model. This is a consequence of the joint locations structure, which still demonstrates a spatial pattern, albeit with reduced intensity, as the mixture of the two samples results in a more evenly distributed set of locations. As a consequence, fitting a preferential model is suitable for the majority of the cases examined.

As a final remark, it is worth noting the importance of evaluating SDM models using out-of-sample data. All our conclusions came from out-of-sample predictive scores (i.e., RMSE and deviation from overall abundance) given that goodness-of-fit scores (i.e., WAIC) did not show significant differences across the three different models.

\section{Conclusion}

Preferential models yield good estimates when combining both preferentialy and randomly collected data to model target species distributions. Preferential models practically perform as well as the more complex mixture model proposed in this study, that accounts for the sampling process of each dataset. Geostatistical models display a tendency for overestimation when the proportion of random samples is low and the complexity of the underlying process is high. However, geostatistical models will also perform well when higher proportions of random samples are available. As well as when there is a large number of samples throughout the study region, and therefore it can be concluded that the area has been sufficiently explored.

%%%%%%%%%%%%%%%%%%%%%%%%%%%%%%%%%%%%%%%%%%%%%%%%%%%%%%

\begin{figure}[h!]
    \centering
    \begin{subfigure}{\linewidth}
        \centering
        \includegraphics[width=.9\linewidth, height=0.45\textheight]{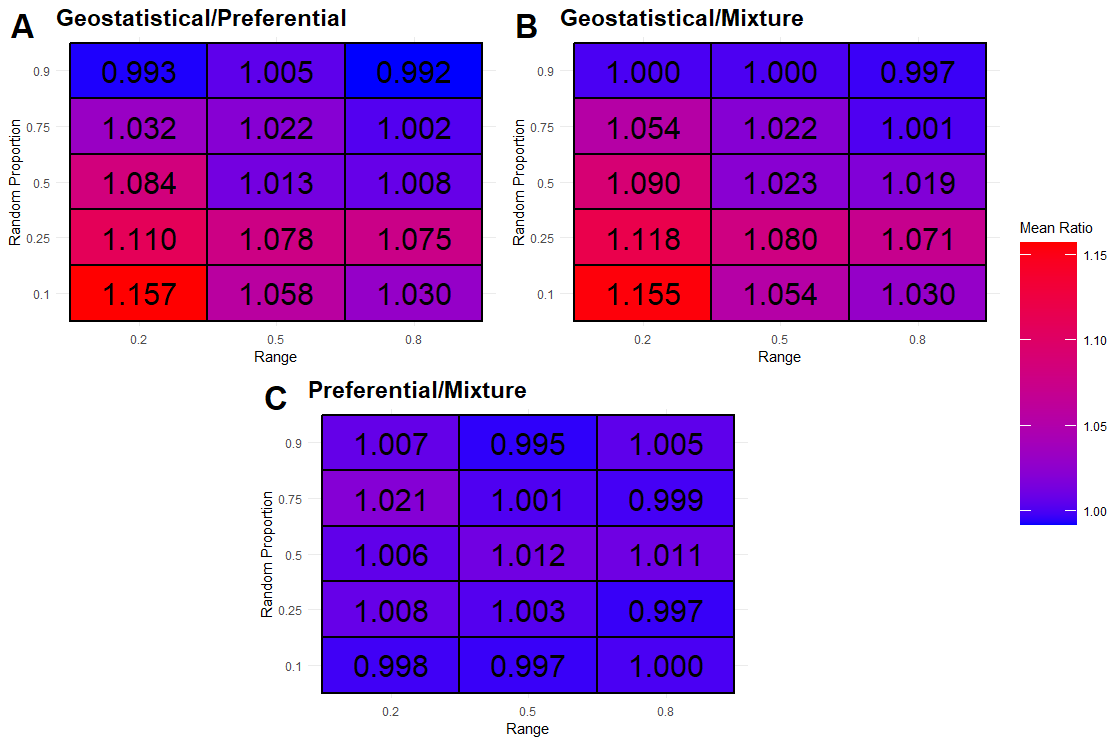}
        \caption{Mean ratio of the RMSE considering spatial range and proportion of random samples jointly.}
        \label{fig:CombinedRMSE}
    \end{subfigure}%
    \vfill%
    \begin{subfigure}{\linewidth}
        \centering
        \includegraphics[width=.9\linewidth, height=0.45\textheight]{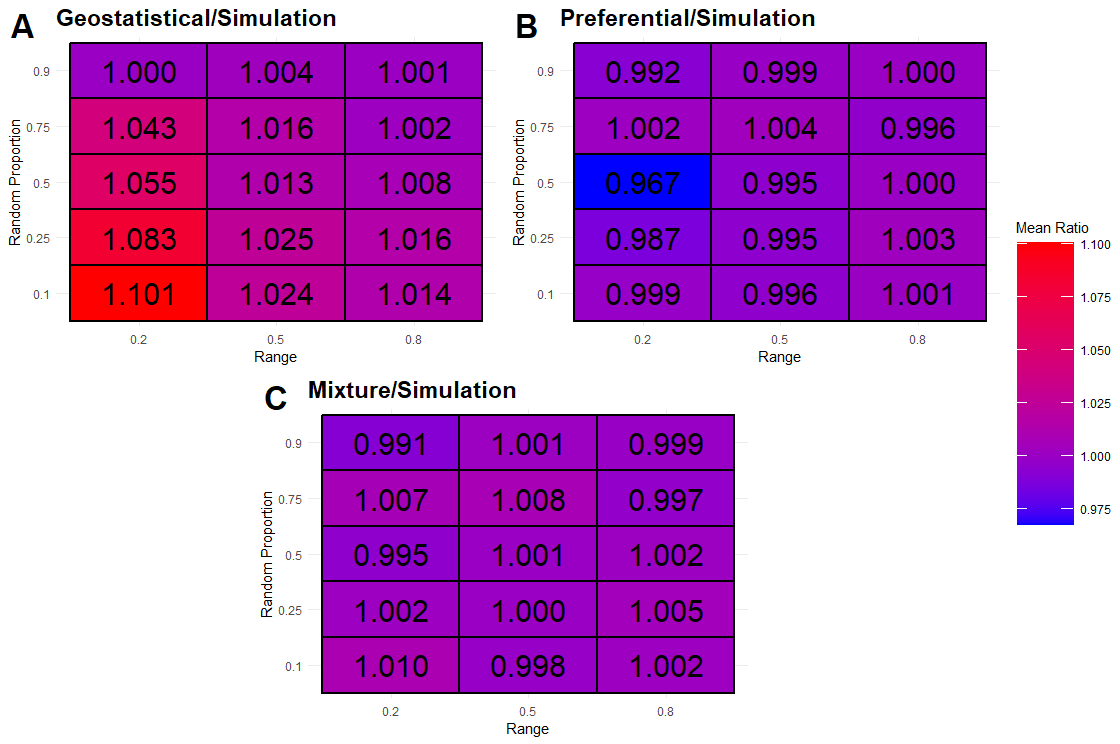}
        \caption{Mean ratio of the total abundance considering spatial range and proportion of random samples jointly.}
        \label{fig:CombinedTotalAbundance}
    \end{subfigure}
    \caption{Mean ratios of the RMSE and total abundance considering the cross values of spatial range and proportion of random samples for each model.}
    \label{fig:CombiendRMSETotalAbundance}
\end{figure}

\cleardoublepage

\bibliographystyle{elsarticle-harv}
\bibliography{article_ipj.bib}

\begin{thebibliography}{11}
\expandafter\ifx\csname natexlab\endcsname\relax\def\natexlab#1{#1}\fi
\providecommand{\url}[1]{\texttt{#1}}
\providecommand{\href}[2]{#2}
\providecommand{\path}[1]{#1}
\providecommand{\DOIprefix}{doi:}
\providecommand{\ArXivprefix}{arXiv:}
\providecommand{\URLprefix}{URL: }
\providecommand{\Pubmedprefix}{pmid:}
\providecommand{\doi}[1]{\href{http://dx.doi.org/#1}{\path{#1}}}
\providecommand{\Pubmed}[1]{\href{pmid:#1}{\path{#1}}}
\providecommand{\bibinfo}[2]{#2}
\ifx\xfnm\relax \def\xfnm[#1]{\unskip,\space#1}\fi
%Type = Article
\bibitem[{Dean et~al.(2022)Dean, El-Shaarawi, Esterby, Mills~Flemming,
  Routledge, Taylor, Woolford, Zidek and Zwiers}]{Preferential_Dean}
\bibinfo{author}{Dean, C.B.}, \bibinfo{author}{El-Shaarawi, A.H.},
  \bibinfo{author}{Esterby, S.R.}, \bibinfo{author}{Mills~Flemming, J.},
  \bibinfo{author}{Routledge, R.D.}, \bibinfo{author}{Taylor, S.W.},
  \bibinfo{author}{Woolford, D.G.}, \bibinfo{author}{Zidek, J.V.},
  \bibinfo{author}{Zwiers, F.W.}, \bibinfo{year}{2022}.
\newblock \bibinfo{title}{Canadian contributions to environmetrics}.
\newblock \bibinfo{journal}{Canadian Journal of Statistics}
  \bibinfo{volume}{50}, \bibinfo{pages}{1355--1386}.
\newblock \DOIprefix\doi{10.1002/cjs.11743}.
%Type = Article
\bibitem[{Diggle et~al.(2010)Diggle, Menezes and
  li~Su}]{GeostatisticalPreferentialSampling_Diggle}
\bibinfo{author}{Diggle, P.J.}, \bibinfo{author}{Menezes, R.},
  \bibinfo{author}{li~Su, T.}, \bibinfo{year}{2010}.
\newblock \bibinfo{title}{Geostatistical inference under preferential
  sampling}.
\newblock \bibinfo{journal}{Journal of the Royal Statistical Society. Series C:
  Applied Statistics} \bibinfo{volume}{59}.
\newblock \DOIprefix\doi{10.1111/j.1467-9876.2009.00701.x}.
%Type = Article
\bibitem[{Diggle et~al.(1998)Diggle, Tawn and Moyeed}]{Geostatistical_Diggle}
\bibinfo{author}{Diggle, P.J.}, \bibinfo{author}{Tawn, J.A.},
  \bibinfo{author}{Moyeed, R.A.}, \bibinfo{year}{1998}.
\newblock \bibinfo{title}{Model-based geostatistics}.
\newblock \bibinfo{journal}{Journal of the Royal Statistical Society. Series C:
  Applied Statistics} \bibinfo{volume}{47}.
\newblock \DOIprefix\doi{10.1111/1467-9876.00113}.
%Type = Book
\bibitem[{Fletcher and
  Fortin(2019)}]{SpatialEcologyConservationModelling_Fletcher}
\bibinfo{author}{Fletcher, R.}, \bibinfo{author}{Fortin, M.J.},
  \bibinfo{year}{2019}.
\newblock \bibinfo{title}{Spatial ecology and conservation modeling:
  Applications with R}.
\newblock \DOIprefix\doi{10.1007/978-3-030-01989-1}.
%Type = Book
\bibitem[{Guisan et~al.(2017)Guisan, Thuiller and
  Zimmermann}]{HabitatSuitability_Guisan}
\bibinfo{author}{Guisan, A.}, \bibinfo{author}{Thuiller, W.},
  \bibinfo{author}{Zimmermann, N.E.}, \bibinfo{year}{2017}.
\newblock \bibinfo{title}{Habitat Suitability and Distribution Models}.
\newblock \bibinfo{publisher}{Cambridge University Press}.
\newblock \DOIprefix\doi{10.1017/9781139028271}.
%Type = Book
\bibitem[{Ovaskainen and Abrego(2020)}]{Joint_Species_Modelling_Ovaskainen}
\bibinfo{author}{Ovaskainen, O.}, \bibinfo{author}{Abrego, N.},
  \bibinfo{year}{2020}.
\newblock \bibinfo{title}{Joint species distribution modelling: with
  applications in R}.
\newblock \bibinfo{publisher}{Cambridge University Press}.
\newblock \DOIprefix\doi{10.1017/9781108591720}.
%Type = Article
\bibitem[{Pennino et~al.(2019)Pennino, Paradinas, Illian, Mu{\~n}oz, Bellido,
  L{\'o}pez-Qu{\'\i}lez and Conesa}]{pennino2019accounting}
\bibinfo{author}{Pennino, M.G.}, \bibinfo{author}{Paradinas, I.},
  \bibinfo{author}{Illian, J.B.}, \bibinfo{author}{Mu{\~n}oz, F.},
  \bibinfo{author}{Bellido, J.M.}, \bibinfo{author}{L{\'o}pez-Qu{\'\i}lez, A.},
  \bibinfo{author}{Conesa, D.}, \bibinfo{year}{2019}.
\newblock \bibinfo{title}{Accounting for preferential sampling in species
  distribution models}.
\newblock \bibinfo{journal}{Ecology and evolution} \bibinfo{volume}{9},
  \bibinfo{pages}{653--663}.
%Type = Article
\bibitem[{Rue et~al.(2009)Rue, Martino and Chopin}]{Rue2009}
\bibinfo{author}{Rue, H.}, \bibinfo{author}{Martino, S.},
  \bibinfo{author}{Chopin, N.}, \bibinfo{year}{2009}.
\newblock \bibinfo{title}{Approximate bayesian inference for latent gaussian
  models by using integrated nested laplace approximations}.
\newblock \bibinfo{journal}{Journal of the Royal Statistical Society. Series B:
  Statistical Methodology} \bibinfo{volume}{71}.
\newblock \DOIprefix\doi{10.1111/j.1467-9868.2008.00700.x}.
%Type = Article
\bibitem[{Rue et~al.(2017)Rue, Riebler, S\o{}rbye, Illian, Simpson and
  Lindgren}]{ReviewINLA_2017}
\bibinfo{author}{Rue, H.}, \bibinfo{author}{Riebler, A.},
  \bibinfo{author}{S\o{}rbye, S.H.}, \bibinfo{author}{Illian, J.B.},
  \bibinfo{author}{Simpson, D.P.}, \bibinfo{author}{Lindgren, F.K.},
  \bibinfo{year}{2017}.
\newblock \bibinfo{title}{Bayesian computing with inla: A review}.
\newblock \bibinfo{journal}{Annual Review of Statistics and Its Application}
  \bibinfo{volume}{4}, \bibinfo{pages}{395--421}.
\newblock \DOIprefix\doi{10.1146/annurev-statistics-060116-054045}.
%Type = Article
\bibitem[{Watanabe(2013)}]{WAIC_Criterion}
\bibinfo{author}{Watanabe, S.}, \bibinfo{year}{2013}.
\newblock \bibinfo{title}{A widely applicable bayesian information criterion}.
\newblock \bibinfo{journal}{Journal of Machine Learning Research}
  \bibinfo{volume}{14}.
%Type = Article
\bibitem[{Watson et~al.(2019)Watson, Zidek and Shaddick}]{Preferential_Watson}
\bibinfo{author}{Watson, J.}, \bibinfo{author}{Zidek, J.V.},
  \bibinfo{author}{Shaddick, G.}, \bibinfo{year}{2019}.
\newblock \bibinfo{title}{{A general theory for preferential sampling in
  environmental networks}}.
\newblock \bibinfo{journal}{The Annals of Applied Statistics}
  \bibinfo{volume}{13}, \bibinfo{pages}{2662 -- 2700}.
\newblock \DOIprefix\doi{10.1214/19-AOAS1288}.

\end{thebibliography}

\cleardoublepage

\end{document}